\documentclass[a4paper,11pt]{article}
\usepackage{jinstpub} 

\proceeding{7$^{\text{th}}$ International Conference on Micro-Pattern Gaseous Detectors\\
11–16 December 2022\\
Rehovot, Israel}

\title{\boldmath Towards robust PICOSEC Micromegas precise timing detectors}

\author[a,b,1]{\footnotesize M. Lisowska\note{Corresponding author.},}
\author[c]{, Y. Angelis}
\author[d]{, S. Aune}
\author[e]{, J. Bortfeldt}
\author[a]{, F. Brunbauer}
\author[c]{, E. Chatzianagnostou}
\author[f]{, K. Dehmelt}
\author[d]{, D. Desforge}
\author[g]{, G. Fanourakis}
\author[a,h]{, K. J. Floethner}
\author[i]{, M. Gallinaro}
\author[j]{, F. Garcia}
\author[f]{, P. Garg}
\author[d]{, I. Giomataris}
\author[k]{, K. Gnanvo}
\author[l]{, T. Gustavsson}
\author[d,2]{, F.J. Iguaz\note{Now at SOLEIL Synchrotron, L’Orme des Merisiers, Départementale 128, 91190 Saint-Aubin, France.}}
\author[a,m,n]{, D. Janssens}
\author[d]{, A. Kallitsopoulou}
\author[o]{, M. Kovacic}
\author[d]{, P. Legou}
\author[p]{, J. Liu}
\author[h,q]{, M. Lupberger}
\author[k]{, S. Malace}
\author[a,c,3]{,I. Maniatis\note{Now at Department of Particle Physics and Astronomy, Weizmann Institute of Science, Hrzl st. 234, Rehovot, 7610001, Israel.}}   
\author[p]{, Y. Meng}
\author[a,q]{, H. Muller}
\author[a]{, E. Oliveri}
\author[a,r]{, G. Orlandini}
\author[d]{, T. Papaevangelou}
\author[s]{, M. Pomorski}
\author[a]{, L. Ropelewski}
\author[c,t]{, D. Sampsonidis}
\author[a,q]{, L. Scharenberg}
\author[a]{, T. Schneider}
\author[s]{, E. Scorsone}
\author[d,4]{, L. Sohl\note{Now at TÜV NORD EnSys GmbH \& Co. KG.}}
\author[a]{, M. van Stenis}
\author[u]{, Y. Tsipolitis}
\author[c,t]{, S.E. Tzamarias}
\author[w]{, A. Utrobicic}
\author[a,x]{, R. Veenhof}
\author[p]{, X. Wang,}
\author[a,y]{, S. White}
\author[p]{, Z. Zhang}
\author[p]{and Y. Zhou}

\affiliation[a]{European Organization for Nuclear Research (CERN), CH-1211, Geneve 23, Switzerland}

\affiliation[b]{Université Paris-Saclay, F-91191 Gif-sur-Yvette, France}

\affiliation[c]{Department of Physics, Aristotle University of Thessaloniki, University Campus, GR-54124, Thessaloniki, Greece}

\affiliation[d]{IRFU, CEA, Université Paris-Saclay, F-91191 Gif-sur-Yvette, France}

\affiliation[e]{Department for Medical Physics, Ludwig Maximilian University of Munich,  Am Coulombwall 1, 85748 Garching, Germany}

\affiliation[f]{Stony Brook University, Dept. of Physics and Astronomy, Stony Brook, NY 11794-3800, USA}

\affiliation[g]{Institute of Nuclear and Particle Physics, NCSR Demokritos, GR-15341 Agia Paraskevi, Attiki, Greece}

\affiliation[h]{Helmholtz-Institut für Strahlen- und Kernphysik, University of Bonn, Nußallee 14–16, 53115 Bonn, Germany}

\affiliation[i]{Laboratório de Instrumentacão e Física Experimental de Partículas, Lisbon, Portugal}

\affiliation[j]{Helsinki Institute of Physics, University of Helsinki, FI-00014 Helsinki, Finland}

\affiliation[k]{Jefferson Lab, 12000 Jefferson Avenue, Newport News, VA 23606, USA}

\affiliation[l]{LIDYL, CEA, CNRS, Universit Paris-Saclay, F-91191 Gif-sur-Yvette, France}

\affiliation[m]{Inter-University Institute for High Energies (IIHE), Belgium}

\affiliation[n]{Vrije Universiteit Brussel, Pleinlaan 2, 1050 Brussels, Belgium}

\affiliation[o]{Faculty of Electrical Engineering and Computing, University of Zagreb, 10000 Zagreb, Croatia}

\affiliation[p]{State Key Laboratory of Particle Detection and Electronics, University of Science and Technology of China, Hefei 230026, China}

\affiliation[q]{Physikalisches Institut, University of Bonn, Nußallee 12, 53115 Bonn, Germany}

\affiliation[r]{Friedrich-Alexander-Universität Erlangen-Nürnberg, Schloßplatz 4, 91054 Erlangen, Germany}

\affiliation[s]{CEA-LIST, Diamond Sensors Laboratory, CEA Saclay, F-91191 Gif-sur-Yvette, France}

\affiliation[t]{Center for Interdisciplinary Research and Innovation (CIRI-AUTH), Thessaloniki 57001, Greece}

\affiliation[u]{National Technical University of Athens, Athens, Greece}

\affiliation[w]{Institute Ruder  Bosković Institute, Bijeni\v{c}ka cesta 54, 10000, Zagreb, Croatia}

\affiliation[x]{Bursa Uludaǧ University, Görükle Kampusu, 16059 Niufer/Bursa, Turkey}

\affiliation[y]{University of Virginia, USA}

\emailAdd{marta.lisowska@cern.ch}

\abstract{

The PICOSEC Micromegas (MM) detector is a precise timing gaseous detector consisting of a~Cherenkov radiator combined with a photocathode and a MM amplifying structure.
A~100-channel non-resistive PICOSEC MM prototype with 10×10 cm$^2$ active area equipped with a~Cesium Iodide (CsI) photocathode demonstrated a time resolution below $\sigma$~=~18~ps.
The objective of this work is to improve the PICOSEC MM detector robustness aspects, i.e. integration of resistive MM and carbon-based photocathodes, while maintaining good time resolution. 
The PICOSEC MM prototypes have been tested in  laboratory conditions and successfully characterised with 150 GeV/c muon beams at the CERN SPS H4 beam line.
The excellent timing performance below $\sigma$~=~20 ps for an individual pad obtained with the 10×10 cm$^2$ area resistive PICOSEC MM of 20 M$\Omega$/$\Box$ showed no significant time resolution degradation as a result of adding a resistive layer.
A single-pad prototype equipped with a 12 nm thick Boron Carbide (B$_4$C) photocathode presented a time resolution below $\sigma$~=~35 ps, opening up new possibilities for detectors with robust photocathodes.
The results made the concept more suitable for the experiments in need of robust detectors with good time resolution.
}


\begin{document}
\maketitle
\flushbottom

\section{Introduction}
\label{sec:intro}

The development of technologies for precise timing detectors has been driven by the demanding environments expected in future High Energy Physics experiments.
The requirements of a~time resolution of tens of picoseconds, stable long-term operation and a large area coverage must be met to make the device suitable. 
Within the PICOSEC Micromegas (MM) collaboration, a~gaseous detector designed to achieve precise timing response is being developed \citep{picoFirstPaper,19pads, NDIP}.
First proof-of-concept single-pad prototypes demonstrated a time resolution below $\sigma$ = 25 ps \citep{picoFirstPaper}.
A recent multipad PICOSEC MM indicated how precise time resolution can be achieved with a non-resistive 100-channel gaseous detector with 10×10 cm$^2$ active area equipped with a CsI photocathode, showing a~time resolution below $\sigma$~=~18~ps \citep{NDIP}.
Although the first tests of alternative, more robust approaches have been performed in the past (i.e. timing measurements of single-pad prototypes with resistive MM and Diamond-Like-Carbon (DLC) photocathodes \citep{LukasPhD}), new developments including the integration of resistive MM in 100-channel modules and robust Boron Carbide (B$_4$C) photocathodes have not been studied up to now.
The objective of this work is to improve the PICOSEC MM detectors robustness aspects while maintaining a good time resolution. 

\section{Detection concept}
\label{sec:2}

\begin{figure}[htbp]
\centering
\includegraphics[width=1\textwidth]{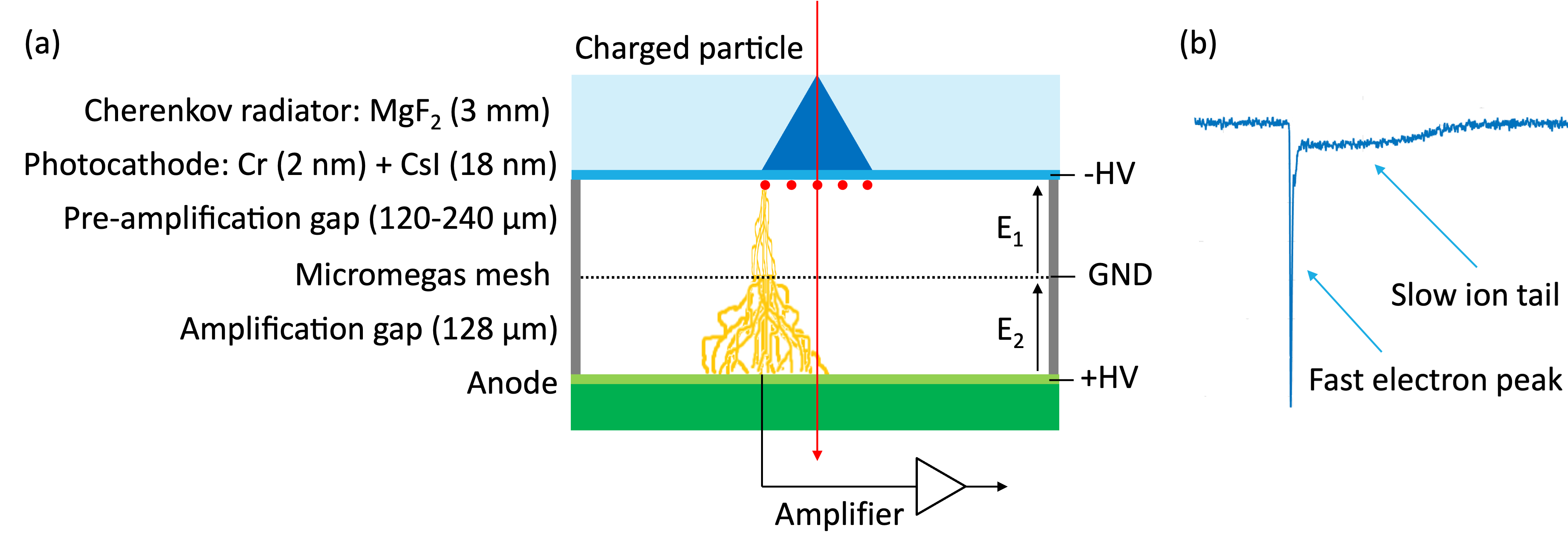}
\caption{(a): PICOSEC MM detection concept. A charged particle passing through a Cherenkov radiator creates UV photons that are converted into electrons on a photocathode, then multiplied in the preamplification gap and afterwards in the amplification gap. (b): A typical PICOSEC MM waveform consisting of a fast electron peak and a slow ion tail.  \label{detectorConcept}}
\end{figure}

The PICOSEC MM detection concept \citep{picoFirstPaper} is illustrated in Fig. \ref{detectorConcept} (a).
A charged particle passing through a Cherenkov radiator, creates a cone of ultraviolet (UV) photons which are converted into  primary electrons on a photocathode coated directly on the radiator. 
Due to a high electric filed, the extracted electrons successively ionise the gas molecules, causing the multiplication of the electrons, first in the preamplification gap, and then, after passing through the MM mesh, in the amplification gap.
A~gas mixture of 80$\%$ Neon, 10$\%$ CF$_4$ and 10$\%$ Ethane at ambient pressure is used to fill both regions.
The amplified electrons induce a signal on the anode which is passed through an amplifier and read out by a digitizer.
A typical PICOSEC MM waveform consists of a~fast electron peak and a slow ion tail, as presented in Fig. \ref{detectorConcept} (b).
The rising edge of the electron peak is used to extract the time of arrival of the charged particle.

\section{Experimental setup}
\label{sec:3}

The PICOSEC MM prototypes have been tested for stable operation in laboratory conditions and successfully characterisied at the CERN SPS H4 beam line with 150 GeV/c muon beams during RD51 test beam campaigns.
The main purpose of the muon beam characterisation was to measure the time resolution of the detectors.
The experimental setup was based on a beam telescope with triggering, timing and tracking capabilities.
Triple-GEMs were used to obtain the precise tracking of the particles, while a~micro-channel plate photomultiplier tube (MCP-PMT, Hamamatsu R3809U50) was used as timing reference (time resolution in the inner part of the active area below $\sigma$ = 4~ps \citep{MCP_PMT}) and data acquisition trigger.
As the new electronics dedicated for multipad device, custom-made RF pulse amplifier cards optimised for PICOSEC \citep{PICOamp} and 128-channel SAMPIC Waveform Time to Digital Converter \citep{SAMPIC} with 8.5~GS/s~sampling frequency were used.

\section{Resistive Micromegas}
\label{sec:4}

Resistive MM are essential to make the PICOSEC concept suitable for physics applications.
The advantages of using resistive MM in the PICOSEC detector include
the limitation of the destructive effect of discharges,
resulting in stable operation under intense particle beams,
as well as a possibly better position reconstruction and signal sharing.
A possible risk of a resistive MM implementation may be to affect the shape of the signal, in particular the rising edge, by spreading the signal due to the resistive layer, which could result in a loss of the timing information.
The objective is to profit from the advantages of the resistive MM while maintaining a good time resolution.
To choose an optimal resistivity, two aspects must be considered.
Firstly, the resistivity must be low enough to minimise the voltage drop during high-rate beam and improve the position reconstruction using the signal weighted average from neighbouring pads.
Secondly, the resistivity needs to be high enough to ensure stable operation and not affect the rising edge of the signal.
Simulations of rate capability and signal rising edge dependence were performed to select the resistivity for the new PICOSEC prototype.
Informed by the simulations, a~100-channel detector with a 10×10 cm$^2$ area resistive MM with anode surface resistivity of 20~M$\Omega$/$\Box$ was produced.
The production procedure was done in the same way as for a non-resistive multipad \citep{NDIP} with an additional step of adding a resistive layer of DLC that was placed on top of the readout pads with an insulating layer in between.

\begin{figure}[htbp]
\centering
\includegraphics[width=.4625\textwidth]{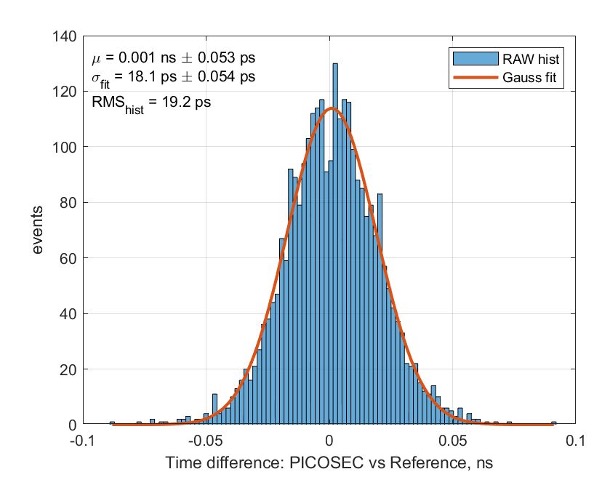}
\caption{SAT distribution of one channel of the resistive PICOSEC MM. The cathode voltage was set at -490~V, the MM mesh at ground and the anode at +290 V. The histogram consists of the data after implementing a geometrical cut of a~4~mm diameter circle around the pad center to include only fully contained events. A~Gaussian fit of the data gives a time resolution of the measured channel of $\sigma$ = 19.2 ps. \label{rMMtiming}}
\end{figure}

First test beam measurements of the multipad equipped with a resistive MM of 20~M$\Omega$/$\Box$, a~CsI photocathode and RF pulse amplifiers were performed with an oscilloscope to compare the results with a non-resistive prototype \citep{NDIP}.
The time resolution of the detector was calculated as the standard deviation of the signal arrival time (SAT) distribution.
The measurements of the 10×10 cm$^2$  resistive PICOSEC MM of 20 M$\Omega$/$\Box$ showed a time resolution below $\sigma$~=~20 ps for an individual pad, as can be seen in Fig. \ref{rMMtiming}.
The results confirm that adding a resistive layer does not affect the time resolution of the detector to a significant extent.

\begin{figure}[htbp]
\centering
\includegraphics[width=0.95\textwidth]{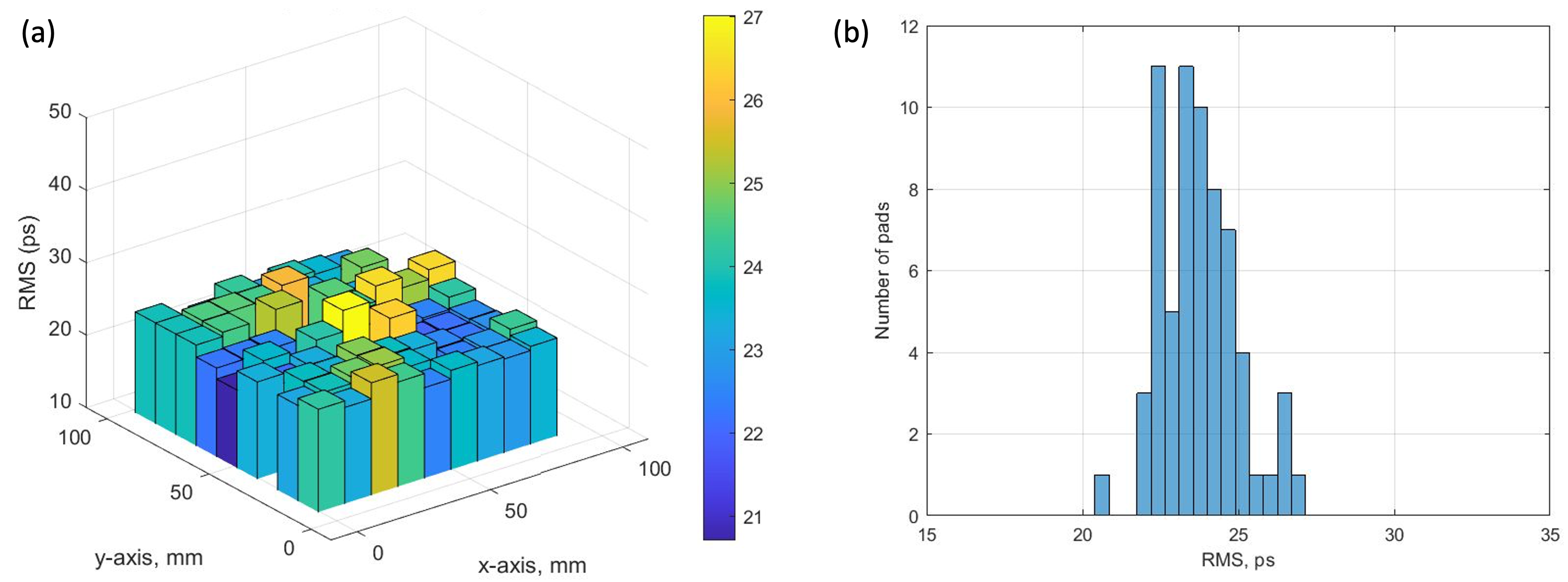}
\caption{Time resolution for the 10×10~cm$^2$ area resistive PICOSEC MM measured using SAMPIC digitiser. Note that only 66 out of 100 pads were measured. (a): 3D plot of the timing response across the channels. (b): Histogram of the time resolution RMS for all measured channels yields the mean value $\sigma\approx$ 23 ps.}
\label{rMMSAMPIC}
\end{figure}

To characterise the full detector, the SAMPIC digitiser was used \citep{SAMPIC}.
The objective was to measure the timing precision.
The measurements of the time resolution across the 10×10~cm$^2$ area resistive PICOSEC MM of 20 M$\Omega$/$\Box$ are presented in Fig. \ref{rMMSAMPIC}.
First results obtained with the new digitizer showed uniform timing response within the pads and a narrow distribution of the time resolution across the channels with the mean value of $\sigma\approx$ 23 ps, proving SAMPIC to be a~suitable tool to study the response of 100-channel PICOSEC detector.

\section{Robust photocathodes}
\label{sec:5}

The base UV-to-electron converter for the PICOSEC MM detector is a semi-transparent CsI photocathode.
The high quantum efficiency of CsI leads to the production of around 10 photoelectrons per minimum ionizing particle (for a 3 mm thick MgF$_2$ radiator with a 3 nm Cr layer and a 18 nm CsI photocathode \citep{picoFirstPaper}). 
However, CsI photocathodes can be easily damaged by ion back flow, sparks, discharges and are sensitive to humidity.
Therefore, there is a need to search for alternative, more robust photocathode materials.
The most promising candidates are B$_4$C, DLC and nanodiamonds.

\begin{figure}[htbp]
\centering
\includegraphics[width=0.95\textwidth]{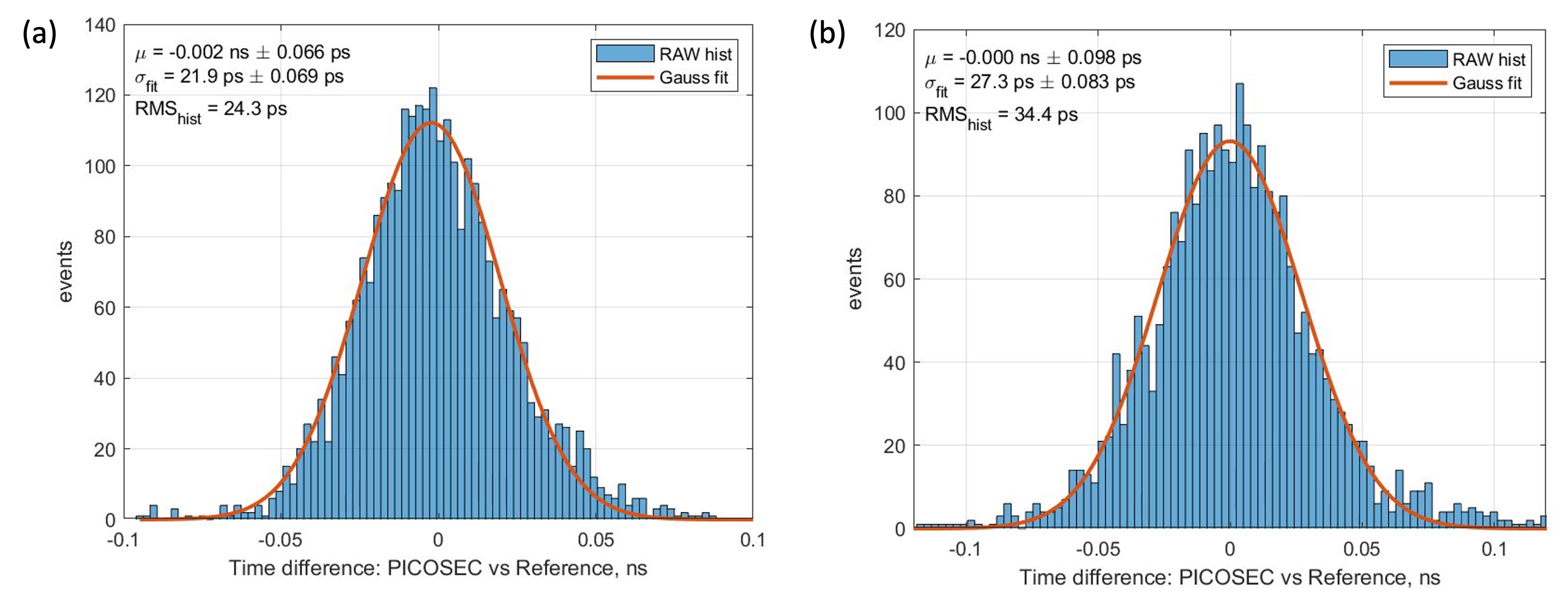}
\caption{SAT distribution of a single-pad prototype equipped with a 12 nm thick B$_4$C photocathode. The voltage on the cathode was set at -490 V and on the anode at +275 V. 
A Gaussian fit yields a time resolution of $\sigma$~=~24.3 ps at the beginning (a) and $\sigma$ = 34.4 ps after 20 hours of operation with gas flushing (b), respectively. \label{B4Ctiming}}
\end{figure}

The measurements of the time resolution for different thicknesses of B$_4$C photocathodes were performed during the RD51 test beam campaigns.
A single-channel non-resistive PICOSEC MM prototype was used to test the samples.
Different photocathode thicknesses ranging from 2 nm to 14 nm were tested.
The best timing response was obtained for a 12 nm thick B$_4$C sample.
First measurements from a single-pad detector with 120 µm preamplification gap equipped with a~12~nm thick B$_4$C photocathode, presented in Fig. \ref{B4Ctiming}, showed a time resolution below $\sigma$~=~25~ps.
The measurements performed 20 hours later presented a time resolution below $\sigma$~=~35~ps.
The 12 nm thick B$_4$C sample demonstrated the best time resolution achieved for robust photocathode materials studied up to now.

\section{Conclusions}
\label{sec:6}

The work described in this paper focused on improving the robustness aspects of the PICOSEC MM detectors, including the integration of resistive MM and robust photocathodes, while maintaining good time resolution.
The excellent timing performance below $\sigma$~=~20 ps for an individual pad obtained with the 10×10 cm$^2$ area resistive PICOSEC MM of 20 M$\Omega$/$\Box$ equipped with a CsI photocathode showed no significant time resolution degradation as a result of adding a resistive layer.
Measurements performed with a scalable readout chain including the SAMPIC digitiser resulted in a successful readout of a multi-channel detector.
A single-pad prototype equipped with a 12 nm thick B$_4$C photocathode presented a time resolution below $\sigma$ = 35 ps, opening up new possibilities for detectors with robust photocathodes.
The results made the concept more suitable for the experiments in need of robust detectors with precise timing.
Developments towards applicable detectors are ongoing.
The challenges include stable operation in intense beams with resistive MM Multipad
and operating a PICOSEC MM detector with a 10×10 cm$^2$ area B$_4$C photocathode.
Scaling up the PICOSEC MM detector by tiling 10×10 cm$^2$ modules or the development of larger prototypes are potential next development steps.


\acknowledgments
We acknowledge the  support of the CERN EP R\&D Strategic Programme on Technologies for Future Experiments; the RD51 collaboration, in the framework of RD51 common projects; the Cross-Disciplinary Program on Instrumentation and Detection of CEA, the French Alternative Energies and Atomic Energy Commission; the PHENIICS Doctoral School Program of Université Paris-Saclay, France; the Program of National Natural Science Foundation of China (grant number 11935014); the COFUND-FP-CERN-2014 program (grant number 665779); the Fundação para a~Ciência e a Tecnologia (FCT), Portugal (CERN/FIS-PAR/0005/2021); the Enhanced Eurotalents program (PCOFUND-GA-2013-600382); the US CMS program under DOE contract No. DE-AC02-07CH11359.




\end{document}